\documentclass[ ]{aastex631}
\usepackage{tabularx}
\usepackage{amsmath}
\usepackage{multirow}
\usepackage{booktabs}
\usepackage{xcolor}  

\usepackage{amsmath}

\begin{document}

\title{MSFA-Net: An Advanced Deep Learning Model for Identifying Blue Horizontal-Branch Stars from LAMOST DR12}

\author{Mingyuan Wang}
\affiliation{School of Airspace Science and Engineering, Shandong University, Weihai 264209, Shandong, People's Republic of China}

\author{Xiaoming Kong}
\affiliation{Shandong Key Laboratory of Space Environment and Exploration Technology, Institute of Space Sciences, School of Space Science and Technology, Shandong University, Weihai 264209, Shandong, People's Republic of China}

\email{xmkong@sdu.edu.cn}

\author{Jie Ju}
\affiliation{School of Sciences, Hebei University of Science and Technology, Shijiazhuang 050018, People's Republic of China}

\author{Yude Bu}
\affiliation{School of Mathematics and Statistics, Shandong University, Weihai 264209, Shandong, People's Republic of China}

\author{Yuchen Liang}
\affiliation{Xi'an University of Architecture and Technology Adelaide University An De College, Xi'an University of Architecture and Technology,  Xi'an, 710311, China}



\begin{abstract}

Blue horizontal-branch (BHB) stars are low-mass, core helium-burning objects with nearly constant luminosities, making them powerful tracers of old, metal-poor populations and valuable standard candles for mapping the Galactic halo. However, robustly identifying BHB stars from low-resolution spectra remains challenging. We present MSFA-Net, a two-stage framework developed for the Large Sky Area Multi-Object Fiber Spectroscopic Telescope (LAMOST)  DR12. By combining multi-scale convolutions with a soft frequency attention mechanism, MSFA-Net learns discriminative representations in both the wavelength domain and the Fourier-frequency domain. On the test set, the framework achieves a precision of 94.67\% in the initial multiclass screening and 98.07\% in the subsequent binary refinement. Applying the trained pipeline to LAMOST DR12, we retrieve 27,853 BHB candidate spectra. After spectral deduplication and removal of previously known objects, we identify 3583 new BHB stars, confirmed via Balmer-line profile fitting. We further estimate atmospheric parameters ($T_{\mathrm{eff}}$, $\log g$, and [Fe/H]) using the machine-learning-based SLAM model and examine their distributions. A non-negligible subset shows unusually high $\log g$ and/or metallicities, which we interpret primarily as inference-related systematics rather than intrinsic properties. Photometric cross-matching with Gaia DR3 and color–magnitude diagrams provide an additional consistency check for the sample. The resulting catalog substantially enlarges the spectroscopically confirmed BHB sample from LAMOST and offers a homogeneous data set for studies of Galactic-halo structure and stellar populations.


\setlength{\parskip}{5pt}

\noindent
\textbf{Unified Astronomy Thesaurus concepts}: Astronomy data analysis (1858); Horizontal branch stars (746); Convolutional neural networks (1938);\\
\textbf{Supporting material}: machine-readable table
\end{abstract}

\section{Introduction}

Blue horizontal-branch (BHB) stars are low-mass, core helium-burning stars surrounded by hydrogen-burning shells \citep{ruhland2011}. In the Hertzsprung–Russell diagram, they occupy the blue side of the RR Lyrae instability strip and trace an old, typically metal-poor stellar population in the Milky Way. Stars with masses of roughly 0.5--1$~M_\odot$ evolve off the red giant branch and populate the blue horizontal branch \citep{montenegro2019,cuplan2021}. Observationally, BHB stars have high and nearly constant luminosities and exhibit mostly A-type (and some late B-type) spectra \citep{barbosa2022,ju2024}. Their atmospheric parameters span effective temperature ($T_\mathrm{eff}$) of approximately 7000--20,000 K, surface gravity ($\log g$) values mainly range from 2.5 to 4.5 dex, and metallicity [Fe/H] is typically below $-1$ \citep{catelan2009,cuplan2021,ju2024}. Compared with main-sequence A-type stars, BHB stars have lower surface gravities and therefore show narrower Balmer-line wings, providing a key spectroscopic diagnostic for classification \citep{vickers2021,barbosa2022}. Moreover, their nearly constant absolute magnitude over a limited color range makes BHB stars valuable standard candles for distance estimation \citep{xue2008}. Owing to these properties, large and reliable BHB samples have become essential for studies of the Milky Way halo, including its structure and kinematics and the mass of the dark matter halo \citep{skiro2004,xue2008}, the detection and characterization of phase-space substructures  \citep{xue2011}, chemo-dynamical correlations and the global halo morphology \citep{bird2022,fuku2024}, and halo density and kinematic profiles \citep{yu2024}.

To obtain high-quality BHB samples, previous studies have mainly adopted two identification strategies: photometric selection and spectroscopic classification. Photometric approaches rely on multi-band photometry and select candidates in color–color or color–magnitude space, exploiting the characteristic colors of BHB stars over a limited range of magnitudes. Using near-infrared photometry from the Two Micron All Sky Survey (2MASS; \citealt{skrutskie2006}), \cite{beers2007} identified 12,056 BHB candidates based on color-index criteria, and  \cite{brown2008} selected 2414 BHB stars by applying constraints in the $JHK$ bands. With optical photometry from the Sloan Digital Sky Survey (SDSS; \citealt{york2000}), \cite{vickers2012} identified 3311 BHB candidates using color–color diagrams. \cite{montenegro2019} subsequently established selection criteria in color space and identified 12,554 BHB stars from the ESO Via Lactea Public Survey. More recently, leveraging the high-precision astrometry and photometry from Gaia, \cite{cuplan2021} combined Gaia EDR3 color and absolute magnitude information \citep{gaia2021} to obtain a much larger sample of 57,377 BHB candidates.

However, photometric selection is sensitive to Galactic extinction and typically suffers from higher contamination than spectroscopic approaches (\citealt{vickers2021,ju2024}). In contrast, spectroscopic methods can more reliably separate BHB stars from photometric look-alikes by exploiting gravity- and temperature-dependent line diagnostics, most notably the Balmer-line profiles and the Ca\,\textsc{ii}\,K feature. Early BHB samples were assembled from narrowband objective-prism surveys \citep{pier1983,beers1988}. Using SDSS spectroscopy, \cite{skiro2004} constructed a sample of 1170 high-Galactic-latitude BHB stars by combining broad-band color cuts with constraints on Balmer-line widths. Building on this framework, \citet{xue2008} and \citet{xue2011} refined the selection criteria and applied them to SDSS DR6 \citep{adelman2008} and SDSS DR8 \citep{aihara2011}, yielding 2558 and 4985 BHB candidates, respectively. More recently, \citet{ju2024} identified 5355 high-confidence BHB stars from LAMOST Data Release 5 \citep{cui2012,zhao2012} by combining line indices with Balmer-line profile fitting. Despite their effectiveness, these traditional pipelines often involve substantial manual tuning and iterative cuts, which becomes increasingly restrictive when confronting the data volume and heterogeneity of modern spectroscopic surveys.

With the rapid development of machine learning and deep learning, automated spectral recognition has become an increasingly important avenue for constructing large BHB samples. Using LAMOST DR5 spectra, \citet{vickers2021} applied an extreme gradient boosting classifier (XGBoost; \citealt{Chen2016}) to compile a catalog of 13,693 BHB candidates. Combining spectroscopy with imaging information, \citet{wei2023} developed a bi-level attention multimodal Transformer (BATMM) based on SDSS DR16 data and identified 4752 BHB candidates. From the Dark Energy Survey DR2 \citep{abbott2021}, \citet{yu2024} constructed a Bayesian mixture model and obtained approximately 2100 candidates. More recently, \citet{zhang2025} proposed a two-stage deep-learning framework, BHBNet, and identified 1605 BHB candidates from LAMOST DR10. Collectively, these efforts highlight the strong potential of data-driven methods---particularly deep neural networks---for scalable and consistent BHB classification across modern sky surveys.

Building on previous studies, we develop a two-stage machine-learning framework to identify BHB stars from low-resolution LAMOST DR12 spectra. In the first stage, we formulate the task as a five-class classification problem and adopt MSFA-Net as the backbone to learn discriminative spectral representations for BHB stars and their main look-alike contaminants. In the second stage, a dedicated binary classifier is employed to further refine the BHB candidates selected from the first stage. We assess the classification performance of each stage in detail and benchmark our approach against several representative machine-learning methods. The trained two-stage classifier is then applied to the full LAMOST DR12 dataset to construct a new BHB catalog, which is further refined through Balmer-line profile fitting.

The paper is organized as follows. Section \ref{sec:data} describes the training and test data sets. Section \ref{sec:method} outlines the MSFA-Net architecture. Section \ref{sec:exp} presents the evaluation metrics, test performance, ablation experiments and comparative experiments. Section \ref{sec:results} analyzes the BHB candidates identified from LAMOST DR12 and investigates their properties. Section \ref{sec:sum} summarizes our main results and discusses prospects for future work.

\section{Data} \label{sec:data}

\subsection{The LAMOST Data}

LAMOST, also known as the Guo Shoujing Telescope, is a 4-m reflecting Schmidt telescope located at Xinglong Observatory in Hebei Province, China, and operated by the Chinese Academy of Sciences. It features a $5^\circ$ field of view and an effective aperture of 3.6--4.9 m. Using 4000 fibers distributed over a focal plane of approximately $20~\mathrm{deg}^2$, LAMOST can obtain spectra for about 4000 objects simultaneously. Its low-resolution spectrograph provides a wavelength coverage of 3800--9000 {\AA} at a resolving power of $R \sim 1800$, reaching a limiting magnitude of $r \sim 18$ mag \citep{cui2012,deng2012,zhao2012}. Since the pilot survey began in October 2011, LAMOST has carried out a series of large-scale spectroscopic surveys and has accumulated tens of millions of spectra, enabling a wide range of studies on Galactic structure and stellar evolution. The most recent Data Release 12 (DR12) contains about 28 million spectra, including approximately 12.6 million low-resolution spectra, and provides stellar parameters for roughly 11.6 million stars.

\subsection{Labeled Data}

This work aims to identify BHB stars from low-resolution LAMOST spectra using a two-stage classification framework. The classifier is not designed as a universal BHB-versus-all classifier for the full LAMOST spectra, but is instead applied to a physically motivated hot-star candidate domain. BHB stars in the temperature range considered here typically exhibit A-type, and occasionally late B-type, spectra with prominent Balmer absorption lines. We therefore first restrict the search space using the spectral classifications provided by the LAMOST pipeline, and then search for BHB stars among A- and B-type spectra with temperatures consistent with the expected BHB regime. Within this candidate domain, the first-stage classifier separates BHB stars from four major classes of hot contaminants, while the second-stage binary classifier further refines ambiguous BHB-like candidates. This design keeps the training and application domains consistent, focusing the model on distinguishing BHB stars from spectroscopically similar hot objects rather than from the full LAMOST population dominated by cooler stars.

For the positive class, we start from the BHB catalog of \citet{ju2025}, in which 5355 BHB stars were identified from LAMOST DR5 low-resolution spectra by combining spectral line indices with Balmer-line profile fitting, and their atmospheric parameters were inferred using a trained deep-learning model. We used the LAMOST observation identifiers (OBSIDs) listed in the catalog of \citet{ju2025} to download the corresponding BHB spectra from LAMOST DR11, resulting in 5176 spectra. To further ensure label reliability and to reduce potential systematic biases in the inferred parameters, we applied astrophysically motivated cuts following \citet{brown2008} and \citet{cuplan2021}. Specifically, we required $[\mathrm{Fe/H}] < -1$, $2.5 \leq \log g \leq 4.5$, and  $7000~\mathrm{K} \leq T_{\mathrm{eff}} \leq 12,000~\mathrm{K}$. After these selections, 4581 high-quality BHB spectra were retained as the final positive sample.  

The negative sample was constructed to represent the main contaminants within the pre-selected A/B-type application domain. It includes normal A- and B-type stars, which occupy the same broad spectral regime as BHB stars and constitute the dominant contaminants in the initial candidate sample. We also included two classes of hot peculiar objects, white dwarfs (WDs) and hot subdwarfs (HSDs), because their blue continua and strong hydrogen absorption lines can resemble those of BHB stars at low spectral resolution and may therefore contaminate the BHB-like candidate space.

Normal A- and B-type spectra were randomly selected from LAMOST DR11 to provide broad and approximately uniform coverage of the relevant atmospheric-parameter space. WD spectra were taken from the catalog of \citet{tan2023}, and HSD spectra were compiled from the catalogs of \citet{lei2018,lei2019,lei2020}, \citet{luo2021}, and \citet{cuplan2022}. These WDs and HSDs were associated with LAMOST DR11 spectra using the LAMOST OBSIDs listed in the source catalogs whenever available. For entries without OBSIDs, we searched for the corresponding LAMOST DR11 spectra by sky position using a matching radius of 2 arcsec. For all classes, we required a signal-to-noise ratio in the $g$ band of $\mathrm{S/N} \geq 10$. This threshold ensures reliable continuum and Balmer-line measurements and defines the selection function of the present search.

\subsection{Data Preprocessing}

The spectra were preprocessed to facilitate stable model training. The procedure consists of three steps:

1. Radial-velocity correction. Each spectrum was shifted to the stellar rest frame using the $z$ value provided in the LAMOST FITS header. For Galactic stars, this value corresponds to the Doppler shift caused by the stellar line-of-sight velocity, rather than a cosmological redshift. The rest-frame wavelength was calculated as:

\begin{equation} 
w_{\rm rest}(n) = \frac{w_{\rm obs}(n)}{1+z}   \label{(1)},
\end{equation}

where $w_{\rm obs}(n)$ is the observed wavelength at pixel $n$, and $z$ is the Doppler-shift reported by the LAMOST pipeline.



2. Wavelength selection and resampling. We restricted the spectra to 4000--7000 Å, retaining the main Balmer absorption lines used for BHB-star identification while avoiding the relatively low-S/N blue end of the LAMOST spectra \citep{li2018}. The selected segment was then resampled onto a common linear wavelength grid with 2430 points using first-order spline interpolation, because the original LAMOST wavelength arrays are not uniformly spaced in linear wavelength. This provides a fixed input dimension while approximately preserving the native sampling density.

3. Min-max scaling. To improve the numerical stability of model optimization, the flux values were rescaled on a per-spectrum basis as

\begin{equation}
    \hat{x} = \frac{x - \min(x)}{\max(x) - \min(x)}, \label{(2)}
\end{equation}

where $x$ denotes the original flux array, and $\min(x)$ and $\max(x)$ are the minimum and maximum flux values within the spectrum, respectively.

\section{Method} \label{sec:method}
\subsection{Model Architecture}

\begin{figure}[ht]  
    \centering  
\includegraphics[width=1\textwidth]{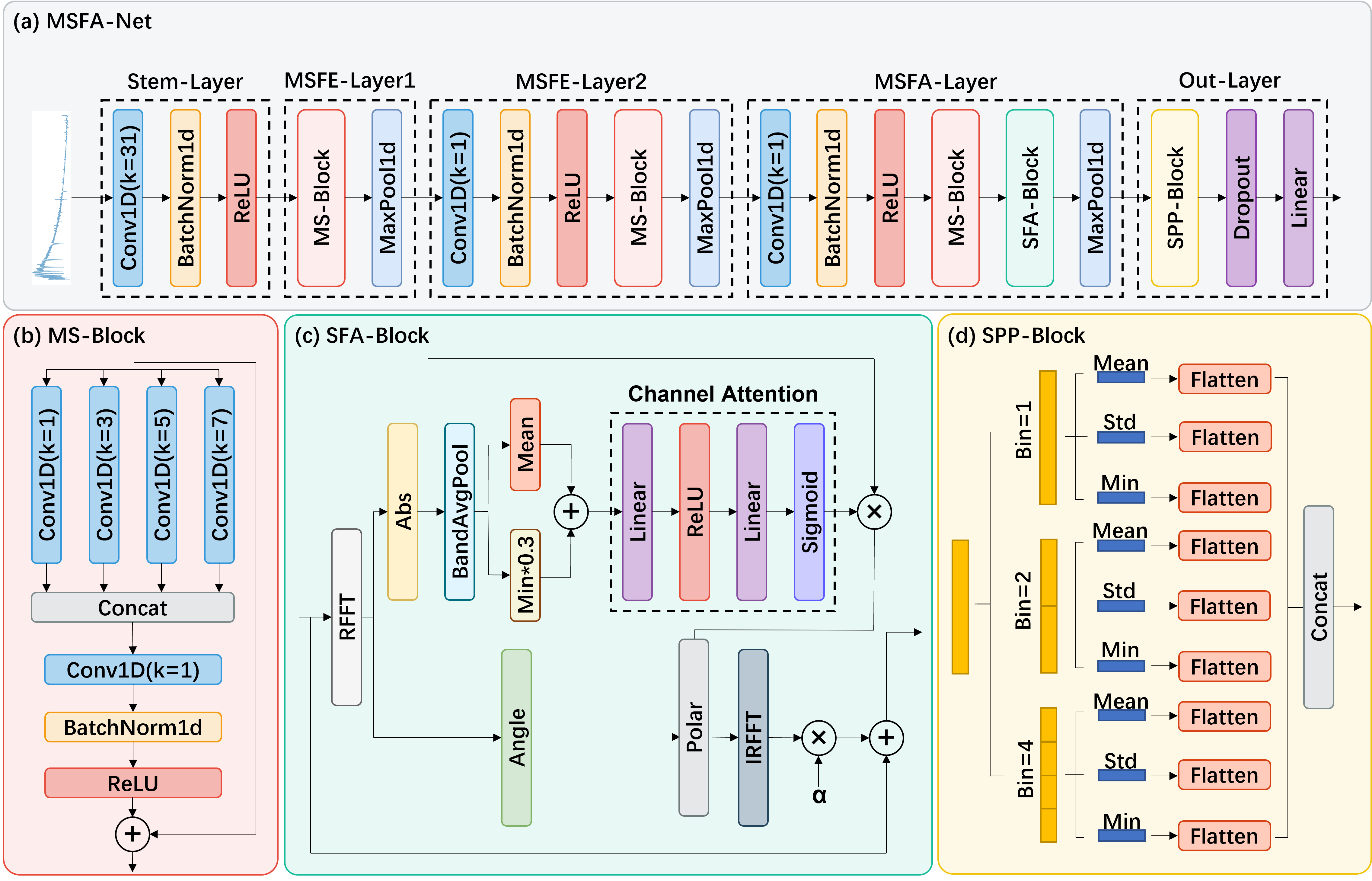}  
    \caption{Hierarchical architecture of the proposed MSFA-Net. (a) Overall framework comprising four sequential stages: Stem, multi-scale spectral feature extraction (MSFE), multi-scale frequency attention (MSFA), and Out-Layer. (b) MS-Block: A multi-scale residual block using parallel one-dimensional convolutions with different kernel sizes to extract multi-granularity spectral features. (c) SFA-Block: A soft frequency-attention block in which intermediate features are transformed into the Fourier domain by the real fast Fourier transform (RFFT). The resulting complex Fourier coefficients are represented by their absolute values, denoted as Abs, and phase angles, denoted as Angle. Channel-wise attention weights are generated from band-wise statistics of the Fourier-amplitude component and applied to this component, while the phase information is preserved. The reweighted amplitude and preserved phase are then reconstructed by the inverse real fast Fourier transform (IRFFT), followed by a learnable residual connection with scaling factor $\alpha$. (d) SPP-Block: A statistical pyramid pooling block that aggregates Mean, standard deviation (Std), and minimum (Min) over hierarchical bin scales of 1, 2, and 4, followed by flattening and concatenation.}
    \label{fig:1}  
\end{figure}

We propose a one-dimensional Multi-Scale Frequency Attention Network (MSFA-Net) to identify BHB stars from low-resolution spectra. A stellar spectrum is treated as an ordered flux sequence along the wavelength axis, in which discriminative patterns may appear at different characteristic scales, such as broad Balmer-line profiles, continuum-shape variations, and relatively narrow absorption features. MSFA-Net is built on a hierarchical multi-scale residual backbone and further introduces Fourier-domain attention into spectral feature learning. In this work, the Fourier-frequency domain refers to the frequency representation of the learned feature sequence along the wavelength axis, rather than the physical electromagnetic frequency of the observed spectrum. By jointly leveraging wavelength-domain and Fourier-domain representations, the network aims to improve class separability for spectra that are difficult to distinguish at low resolution.

As shown in Figure~\ref{fig:1}, MSFA-Net comprises four stages: a spectral feature embedding stage, denoted as the Stem-Layer; two multi-scale spectral feature extraction stages, denoted as MSFE-Layer1 and MSFE-Layer2; a multi-scale frequency-attention stage, denoted as the MSFA-Layer; and an output stage, denoted as the Out-Layer, for feature aggregation and classification. The architecture incorporates three key building blocks: a multi-scale convolution residual block, namely the MS-Block, with parallel kernels \ensuremath{k=\{1,3,5,7\}} for multi-granularity spectral pattern mining; an FFT-based soft frequency-attention block \citep{sfa}, namely the SFA-Block, for frequency-aware feature refinement; and a statistical pyramid pooling block \citep{he2014}, namely the SPP-Block, for robust fixed-length feature aggregation.

The Stem-Layer follows a standard one-dimensional spectral embedding design. It consists of a one-dimensional convolution with a large kernel size of 31, followed by batch normalization (BN; \citealt{ioffe2015}) and a ReLU activation \citep{glorot}. The large kernel provides an initial receptive field along the wavelength axis and maps the raw flux sequence into stable low-level feature representations for subsequent modeling.

MSFE-Layer1 is designed to extract shallow multi-granularity spectral features. It applies an MS-Block followed by max pooling. The MS-Block uses parallel one-dimensional convolutional branches with kernel sizes of 1, 3, 5, and 7, enabling the network to capture local patterns with different characteristic widths, such as relatively narrow metal absorption features and broader Balmer-line profiles. The outputs of these branches are concatenated and fused by a pointwise convolution, followed by residual aggregation and nonlinear activation. The subsequent max-pooling operation downsamples the feature sequence along the wavelength axis and facilitates hierarchical feature abstraction.

MSFE-Layer2 further deepens the feature hierarchy and produces more discriminative intermediate representations. It first uses a channel-expanding pointwise convolution, followed by BN and ReLU, to increase the representational capacity of the feature maps. A second MS-Block is then applied to refine the multi-scale spectral patterns at a higher semantic level, and max pooling is used to further reduce the sequence length. This stage bridges the shallow wavelength-domain features and the subsequent dual-domain modeling in the MSFA-Layer.

The MSFA-Layer serves as the core frequency-aware stage of MSFA-Net, refining high-level features across both the wavelength and Fourier-frequency domains. It begins with a channel-expanding pointwise convolution, followed by an MS-Block for local multi-scale spectral modeling. The resulting features are then processed by the SFA-Block for complementary frequency-domain modulation and subsequently downsampled. Through this sequential dual-domain refinement, the MSFA-Layer enhances the discriminative representation of low-resolution spectra.

Within the SFA-Block, the intermediate feature maps are transformed along the wavelength axis into the Fourier domain to provide a frequency-domain view of the learned spectral representation. This operation complements the local patterns extracted in the wavelength domain by revealing how the feature sequence is distributed over different frequency components. For a real-valued sequence \ensuremath{x_n} with \ensuremath{n=0,\ldots,L-1}, the discrete Fourier transform is defined as

\begin{equation}
\hat{x}_k =
\sum_{n=0}^{L-1}
x_n
\exp\left(
-\frac{2\pi i}{L}nk
\right),
\quad
k=0,\ldots,L-1 ,
\end{equation}

Here, $x_n$ denotes the feature value at the $n$-th wavelength position, $L$ is the sequence length along the wavelength axis, and $\hat{x}_k$ is the complex Fourier coefficient corresponding to the $k$-th frequency component. The exponential term defines the Fourier basis, which projects the wavelength-domain sequence onto sinusoidal components of different frequencies. In this way, the transform decomposes the learned spectral representation into frequency components, where low-frequency responses describe slowly varying spectral trends and high-frequency responses capture rapid local variations.

Since the learned feature maps are real-valued, the real fast Fourier transform is adopted to obtain a compact non-redundant spectrum. The resulting complex spectrum can be interpreted through its Fourier amplitude and phase components. The amplitude reflects the strength of different frequency components and is therefore suitable for adaptive modulation, whereas the phase carries structural information related to the relative arrangement of patterns along the wavelength axis.

Based on this observation, the SFA-Block applies soft frequency attention to the amplitude spectrum while preserving the original phase. The amplitude responses are summarized over coarse frequency bands to form channel-wise frequency descriptors, which are then used to generate attention weights for adaptively rescaling the amplitude spectrum. Here, the term soft indicates that the module does not explicitly select or discard fixed frequency bands; instead, it learns continuous reweighting factors according to the input representation.

After frequency-domain modulation, the adjusted amplitude is recombined with the unchanged phase to reconstruct a modified complex spectrum, which is then mapped back to the wavelength domain using the inverse real Fourier transform. The reconstructed representation is integrated with the original input through a learnable residual connection. This design allows the network to incorporate frequency-domain refinement while preserving the positional structure and optimization stability of the wavelength-domain features.

The Out-Layer consists of an SPP-Block, followed by a dropout layer and a fully connected layer. The SPP-Block aggregates high-level features through multi-scale adaptive pooling with bin sizes of 1, 2, and 4 along the wavelength axis. For each channel and each bin, the mean, standard deviation, and minimum values are computed across positions within the bin. The minimum statistic is particularly effective at capturing the depth of absorption features, complementing mean and standard-deviation-based summaries. These statistics are concatenated to form a compact fixed-length feature representation. Dropout is applied to mitigate overfitting, and the final fully connected layer maps the aggregated features to the classification scores.

\begin{deluxetable}{@{}l c c@{}}
    \setlength{\tabcolsep}{6mm}
    \tablecaption{Output Parameters of Each Layer in MSFA-Net\label{tab:1}}
    \tablewidth{0pt}
    \tablehead{
        \multicolumn{1}{@{}l}{No.} & 
        \colhead{Layer} & 
        \multicolumn{1}{c@{}}{Output Size}
    }
    \startdata
    0 & Input        & $1 \times 2430$ \\
    1 & Stem-Layer   & $64 \times 2430$ \\
    2 & MSFE-Layer1  & $64 \times 1215$ \\
    3 & MSFE-Layer2  & $128 \times 607$ \\
    4 & MSFA-Layer   & $256 \times 303$ \\
    5 & Out-Layer    & $1 \times 5$ or $1 \times 2$ \\
    \enddata
\end{deluxetable}

\subsection{Loss Function}

To improve the robustness of BHB-star identification, we adopted a Weighted label-smoothing cross-entropy (W-LSCE) loss \citep{sze2015}. Label smoothing replaces the hard one-hot target used in the standard cross-entropy loss with a softened target distribution, thereby reducing over-confident predictions, alleviating poor probability calibration, and improving generalization on noisy or ambiguous spectra. Specifically, the label-smoothed target probability $q_{i}(k)$ for class $k$ is defined as:

\begin{equation}
q_{i}(k) =
\begin{cases}
1 - \varepsilon & k = y_{i} \\
\dfrac{\varepsilon}{K - 1} & k \neq y_{i}
\end{cases},
\label{eq:smooth_target}
\end{equation}

where $y_i$ is the ground-truth class label, $K$ is the number of classes, and $\varepsilon$ is the smoothing factor. We used $K=5$ in the first stage and $K=2$ in the second stage, with $\varepsilon=0.1$ in both cases. 

In addition to target smoothing, we introduced class-specific weights to mitigate the class imbalance in the second-stage binary classification. In this stage, the ratio of positive to negative samples is approximately 1:1.7. To reduce the bias toward the majority class, we employed a cost-sensitive formulation by incorporating class-specific weights $w \in \mathbb{R}^K$ into the cross-entropy loss. The weight for class $k$ was defined as

\begin{equation}
w_k = \frac{1}{\ln(N_k + 2)},
\label{eq:class_weight}
\end{equation}

 where \ensuremath{N_k} is the number of training samples in class \ensuremath{k}. This logarithmic inverse-frequency weighting assigns relatively larger weights to classes with fewer samples, while avoiding excessively large weights for rare classes. By combining the smoothed targets \ensuremath{q_i(k)} with the class weights \ensuremath{w_k}, the final W-LSCE loss is formulated as

\begin{equation}
\mathcal{L} = - \frac{1}{N} \sum_{i=1}^{N} \sum_{k=1}^{K} w_{k}\, q_{i}(k)\, \log\!\big(p_{i}(k)\big),
\label{eq:loss_final}
\end{equation}

 where \ensuremath{N} is the number of training samples, $q_i(k)$ is the label-smoothed target distribution defined in Eq.~\eqref{eq:smooth_target}, and $p_i(k)$ is the predicted probability of assigning sample $i$ to class $k$. This integrated weighting and smoothing scheme simultaneously improves probability calibration and compensates for the moderate class imbalance in the second-stage classification, thereby increasing the penalty for misclassifying the minority BHB class and helping the model capture the subtle spectral features required for reliable separation from spectroscopically similar contaminants. We applied the W-LSCE loss in both classification stages for methodological consistency. In the first-stage classification, where no substantial class imbalance is present, the class weights are nearly uniform and therefore mainly act as an approximately constant scaling factor, rather than introducing additional class-dependent bias.

\subsection{Model Optimization}

To facilitate stable convergence and enhance discriminative performance for BHB star identification, the network was trained using the AdamW optimizer \citep{loshchilov2017} with an initial learning rate of $1\times10^{-3}$. The learning rate was adaptively decayed by a factor of 0.5 if the validation metric showed no improvement for 8 consecutive epochs,  with a minimum threshold of $1\times10^{-6}$. Training and testing were conducted with a batch size of 96.

\section{Experiment} \label{sec:exp}
\subsection{Evaluation Metrics}

The objective of this study is to discriminate BHB stars from non-BHB stars. To comprehensively evaluate the classification performance, we report four widely used metrics: accuracy, precision, recall, and F1\_score. These metrics capture complementary aspects of model behavior, including overall correctness, reliability of positive predictions, sensitivity to true BHB stars, and the balance between precision and recall.

\begin{equation}  
\mathrm{Accuracy} = \frac{\mathrm{TP} + \mathrm{TN}}{\mathrm{TP} + \mathrm{FP} + \mathrm{TN} + \mathrm{FN}},
\label{eq:acc}
\end{equation}

\begin{equation}  
\mathrm{Precision} = \frac{\mathrm{TP}}{\mathrm{TP} + \mathrm{FP}},
\label{eq:4}
\end{equation}

\begin{equation}  
\mathrm{Recall} = \frac{\mathrm{TP}}{\mathrm{TP} + \mathrm{FN}},
\label{eq:5}
\end{equation}

\begin{equation}  
\mathrm{F1\_score} = \frac{2 \times \mathrm{Precision} \times \mathrm{Recall}}{\mathrm{Precision} + \mathrm{Recall}},
\label{eq:6}
\end{equation}

In these equations, $\mathrm{TP}$ (true positives) denotes the number of BHB stars correctly classified as BHB, while $\mathrm{TN}$ (true negatives) denotes the number of non-BHB stars correctly classified as non-BHB. Conversely, $\mathrm{FP}$ (false positives) corresponds to non-BHB stars misclassified as BHB, and $\mathrm{FN}$ (false negatives) corresponds to BHB stars misclassified as non-BHB. Accordingly, precision quantifies the proportion of true BHB stars among all instances predicted as BHB, recall measures the fraction of BHB stars that are correctly identified, and the F1\_score is the harmonic mean of precision and recall.

\subsection{Classification Construction}

\begin{deluxetable}{@{}l c c c@{}} 
    \setlength{\tabcolsep}{5mm} 
    \tablecaption{Dataset Categories and Quantities\label{tab:dataset}}
    \tablewidth{0pt}
    \tablehead{
        \multicolumn{1}{@{}l}{Stage} & 
        \colhead{Class Type} &  
        \colhead{Class} & 
        \multicolumn{1}{c@{}}{Number}
    }
    \startdata
    \multirow{5}{*}{1}
    & \multirow{5}{*}{Multi-class} 
    & BHB               & 4581 \\
    &                   & A-type star       & 4500 \\
    &                   & B-type star       & 4500 \\
    &                   & White dwarf       & 4500 \\
    &                   & Hot subdwarf star & 3431 \\
    \hline
    \multirow{3}{*}{2}
    & Positive          
    & BHB               & 4581 \\
    & \multirow{2}{*}{Negative} 
    & A-type star       & 4500 \\
    &                   & Hot subdwarf star & 3431 \\
    \enddata
\end{deluxetable}

\begin{deluxetable}{@{}l c c c@{}}
    \setlength{\tabcolsep}{6mm}
    \tablecaption{Results of the Five-class Classification Model\label{tab:performance_stage1}}
    \tablewidth{0pt}
    \tablehead{
        \multicolumn{1}{@{}l}{Class} & 
        \colhead{Precision (\%)} & 
        \colhead{Recall (\%)} & 
        \multicolumn{1}{c@{}}{F1\_score (\%)}
    }
    \startdata
    BHB star           & 94.67 & 98.80 & 96.69 \\
    A-type star        & 92.77 & 90.36 & 91.55 \\
    B-type star        & 95.22 & 92.62 & 93.90 \\
    Hot subdwarf star  & 86.91 & 91.74 & 89.26 \\
    White dwarf        & 93.14 & 90.10 & 91.60 \\
    \hline
    Total accuracy (\%) & \multicolumn{3}{c}{92.80} \\
    \enddata
\end{deluxetable}

This section describes the construction of a two-stage classification framework for BHB star identification. In the first stage, a five-class model is trained to distinguish BHB stars from four non-BHB stellar types: A-type stars, B-type stars, HSDs, and WDs. Given the vast volume and the great spectral diversity of the target dataset, the first-stage model inevitably yields a candidate sample with residual contamination. To further purify the BHB candidates, a second stage is introduced. Analysis of the misclassified samples from the first-stage results reveals that A-type stars and HSDs are the primary contaminants. Consequently, we design a binary classifier that specifically focuses on discriminating BHB stars from these two spectroscopically similar classes. The sample composition and class distributions for each stage are summarized in Table~\ref{tab:dataset}. For both stages, the dataset is randomly partitioned into training and test subsets with an 8:2 ratio.

In the first stage, the five-class model was trained following the optimization and evaluation protocols described above. On the test set, as shown in Table~\ref{tab:performance_stage1}, it achieved an overall accuracy of 92.80\%. For the target BHB class, the precision, recall, and F1\_score reached 94.67\%, 98.80\%, and 96.69\%, respectively. In the second stage, we fine-tuned the network using the parameters initialized from the first-stage model, updating all weights (i.e., full-parameter transfer) to adapt the classifier from the five-class task to the binary BHB versus non-BHB task. For the BHB class, the precision, recall, and F1\_score reached 98.07\%, 98.49\%, and 98.28\%, respectively.

\subsection{Ablation Experiments}

Ablation experiments were conducted to evaluate the effects of the pooling strategy and the placement of the SFA block. As summarized in Table~\ref{tab:ablation_exp}, replacing global average pooling (GAP) with SPP generally improves performance, particularly in the binary classification stage and in precision for the five-class stage. To examine the impact of frequency-domain attention, the SFA block was separately embedded in the MSFE-Layer2 or the MSFA-Layer positions. While introducing SFA at the MSFE-Layer2 position yields clear performance gains, positioning it at the MSFA-Layer position achieves the best results, producing the highest F1\_scores in both stages. Accordingly, the MSFA-Layer configuration is adopted as the final MSFA-Net architecture.

\begin{deluxetable*}{@{}l c c c c@{}}
    \setlength{\tabcolsep}{6mm}
    \tablecaption{Ablation performance of the proposed method for the five-class and binary classification tasks.\label{tab:ablation_exp}}
    \tablewidth{0pt}
    \tablehead{
        \multicolumn{1}{@{}l}{\multirow{2}{*}{Algorithm}} & 
        \multicolumn{2}{c}{Five-class classification} & 
        \multicolumn{2}{c@{}}{Binary classification} \\
        \cline{2-3} \cline{4-5}
        \colhead{} & 
        \colhead{Precision (\%)} & 
        \colhead{F1\_score (\%)} & 
        \colhead{Precision (\%)} & 
        \multicolumn{1}{c@{}}{F1\_score (\%)}
    }
    \startdata
    MS-ResNet+GAP         & 94.97 & 94.12 & 95.76 & 96.18 \\
    MS-ResNet+SPP         & 95.02 & 94.05 & 96.52 & 96.60 \\
    MSFA-Net (Layer2 SFA) & 96.62 & 94.13 & 97.13 & 97.45 \\
    MSFA-Net              & 94.67 & 96.69 & 98.07 & 98.28 \\
    \enddata
\end{deluxetable*}

\subsection{Comparative Experiments}

A comparative analysis was conducted to evaluate the discrimination capability of the proposed MSFA-Net against representative baseline models, including VGG~\citep{simonyan2015}, AlexNet~\citep{alex}, ResNet18~\citep{he2016}, a baseline CNN, and BHBNet~\citep{zhang2025}. All models are trained and evaluated using the same data partitioning and training protocol to ensure a fair comparison. As summarized in Table~\ref{tab:comp_exp}, MSFA-Net consistently outperforms these baselines in terms of precision, recall, and F1\_score.

\begin{deluxetable*}{@{}l c c c c@{}}
    \setlength{\tabcolsep}{6mm}
    \tablecaption{Comparative performance of MSFA-Net and several reference models across two classification stages.\label{tab:comp_exp}}
    \tablewidth{0pt}
    \tablehead{
        \multicolumn{1}{@{}l}{\multirow{2}{*}{Algorithm}} & 
        \multicolumn{2}{c}{Five-class classification} & 
        \multicolumn{2}{c@{}}{Binary classification} \\
        \cline{2-3} \cline{4-5}
        \colhead{} & 
        \colhead{Precision (\%)} & 
        \colhead{F1\_score (\%)} & 
        \colhead{Precision (\%)} & 
        \multicolumn{1}{c@{}}{F1\_score (\%)}
    }
    \startdata
    CNN      & 91.38 & 90.78 & 91.14 & 93.05 \\
    AlexNet  & 93.51 & 93.92 & 94.77 & 94.89 \\
    VGG11    & 93.62 & 93.21 & 95.30 & 95.75 \\
    ResNet18 & 94.04 & 93.67 & 95.64 & 96.06 \\
    BHBNet   & 93.48 & 93.87 & 96.01 & 96.18 \\
    MSFA-Net & 94.67 & 96.69 & 98.07 & 98.28 \\
    \enddata
\end{deluxetable*}

\section{Results and Discussion} \label{sec:results}
\subsection{Large-scale Search for BHB Candidates}

We searched for BHB stars in the LAMOST DR12 low-resolution spectra by selecting A- and B-type stars with $g$-band S/N $\geq 10$, resulting in 597,990 spectra. After applying the same preprocessing pipeline as used in model training, the spectra were first classified using the proposed two-stage MSFA-Net framework. In the first stage, the five-class model identified 37,441 BHB candidate spectra. These candidate spectra were further refined by the second-stage binary classifier with a BHB confidence threshold of 0.5, yielding 27,853 BHB candidate spectra. To identify newly discovered BHB stars, we cross-matched these candidates with existing catalogs \citep{xue2011,vickers2021,ju2025,zhang2025} using TOPCAT and removed previously reported objects. Specifically, for the catalogs of \cite{ju2025} and \cite{zhang2025}, we performed the cross-match based on LAMOST OBSIDs. For \cite{xue2011} and \cite{vickers2021}, which lack OBSIDs, we cross-matched by sky position using a matching Max Error of 3 arcsec. After this de-duplication process, we identified 16,721 newly discovered BHB candidate spectra, corresponding to 15,015 stars.

\subsection{Balmer Line Profile Fitting}

To further refine the MSFA-Net candidates, we employed the $D_{0.2}$--$f_m$ diagnostic \citep{pier1983,yanny2000}. Here, $D_{0.2}$ denotes the Balmer-line width measured at 20\% below the normalized continuum, and $f_m$ represents the line-core flux relative to the local continuum. Since Balmer-line shapes are mainly governed by effective temperature and surface gravity \citep{brown2008}, resulting in prominent yet comparatively narrow profiles for BHB stars, this plane provides an efficient means to separate them from common contaminants such as blue stragglers (BS), A-type main-sequence (MS) stars, and hot subdwarfs \citep{ju2024}.

\begin{figure}[ht]
    \centering
    \begin{minipage}{0.42\textwidth}
        \centering
        \includegraphics[width=\textwidth]{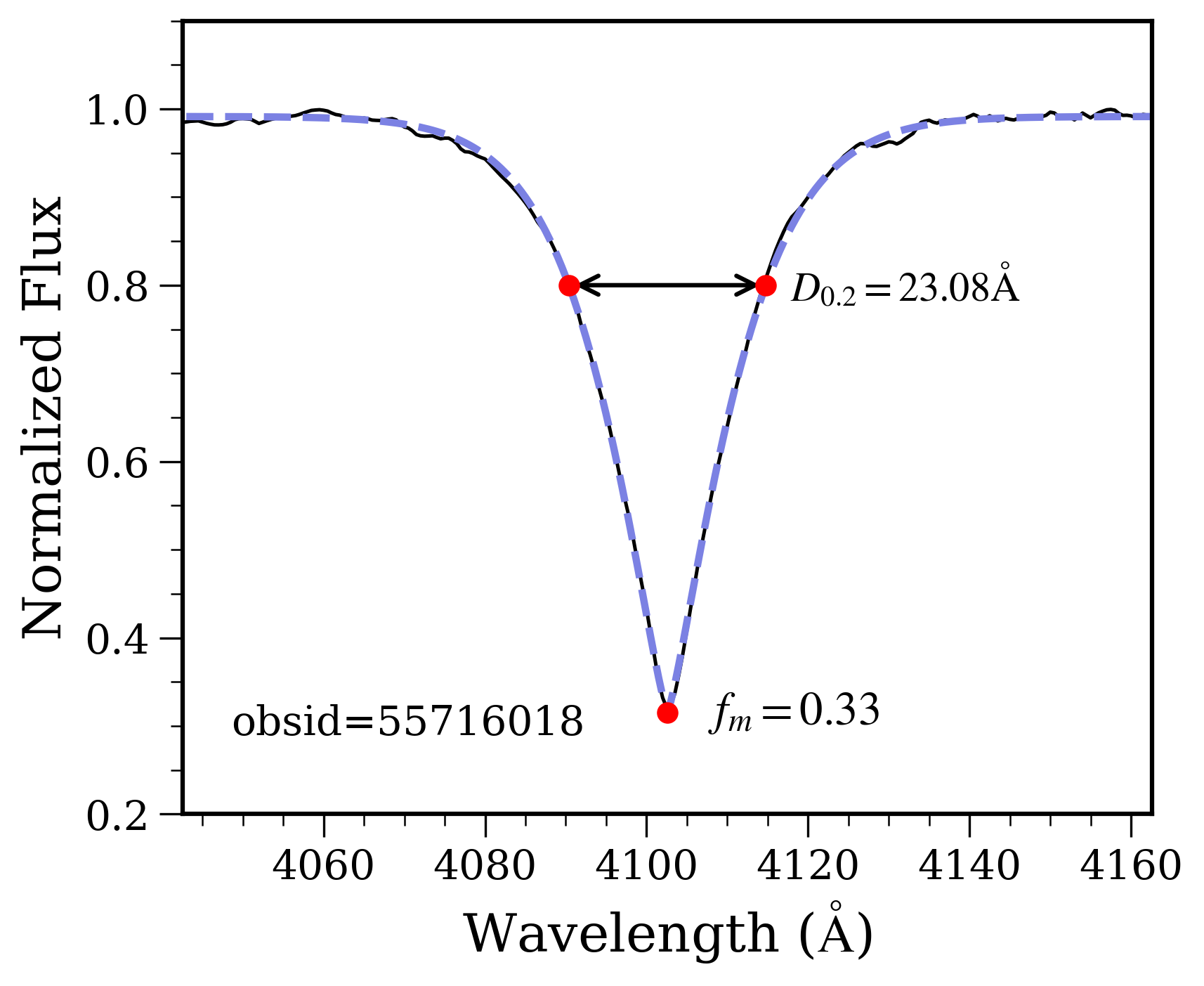}
    \end{minipage}%
    \begin{minipage}{0.42\textwidth}
        \centering
        \includegraphics[width=\textwidth]{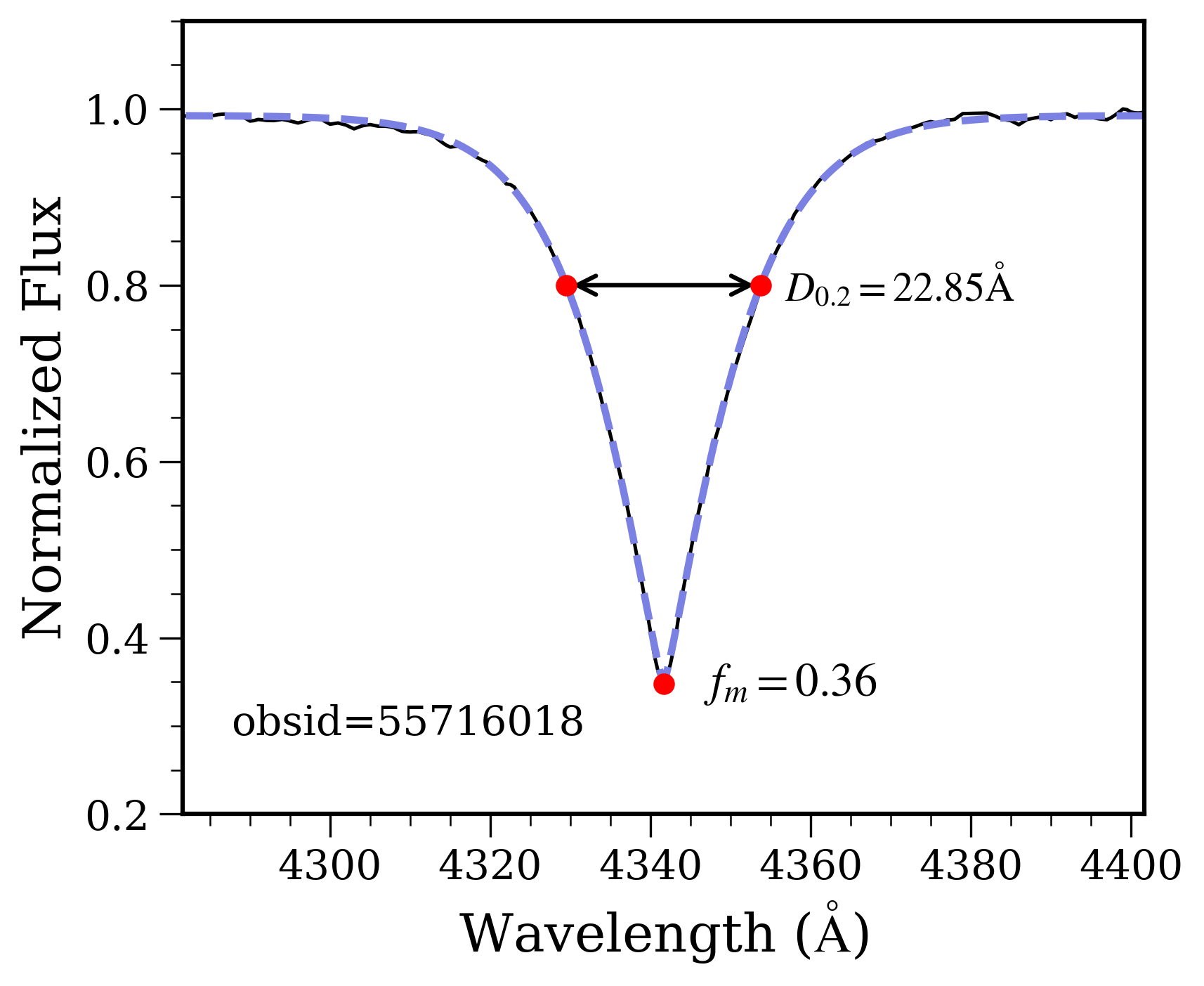}
    \end{minipage}
    \caption{Normalized spectra near H$_{\delta}$ (left) and H$_{\gamma}$ (right) for a BHB candidate (obsid = 55716018). The black solid curve shows the normalized spectrum, and the blue dashed curve indicates the best-fitting SWS profile. The values of $D_{0.2}$ and $f_m$ are marked.}
    \label{fig:fit_plot}
\end{figure}

We modeled the Balmer-line profiles using the scale--width--shape (SWS) parametrization \citep{clewley2002,clewley2004}. The analysis was restricted to H\ensuremath{_{\delta}} and H\ensuremath{_{\gamma}}, because the higher-order Balmer lines are increasingly crowded, making reliable continuum placement difficult \citep{clewley2002}. Before fitting the line profiles, we performed a local continuum normalization around each selected Balmer line. To account for small residual errors in radial-velocity correction and local continuum normalization, we adopted the five-parameter S\'{e}rsic profile \citep{xue2008}:

\begin{equation}
y = n - a \exp\left[-\left(\frac{|\lambda - \lambda_{0}|}{b}\right)^{c}\right],
\label{eq:sws_fit}
\end{equation}

where $y$ is the normalized flux at wavelength $\lambda$, $\lambda_{0}$ is the nominal central wavelength of the Balmer line, and $n$ allows for a local continuum offset. The parameters $a$, $b$, and $c$ control the line depth, characteristic width, and profile shape, respectively. Figure~\ref{fig:fit_plot} illustrates the best-fitting SWS profiles for a representative candidate. The diagnostic parameters $D_{0.2}$ and $f_m$ were derived directly from these best-fitting models.

\begin{figure}[ht]
    \centering
    \begin{minipage}{0.42\textwidth}  
        \centering
        \includegraphics[width=\textwidth]{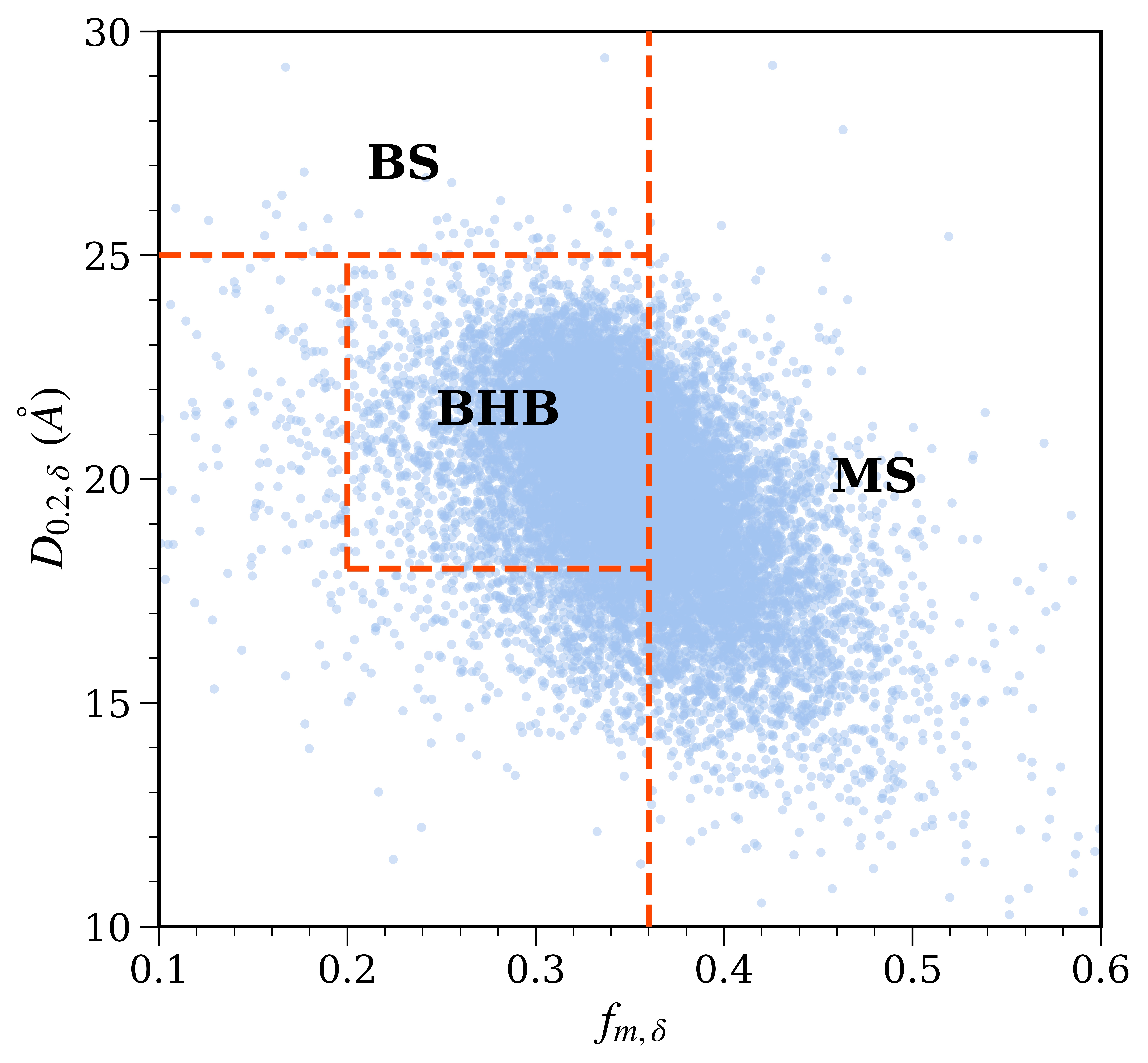} 
    \end{minipage}
    \begin{minipage}{0.42\textwidth}
        \centering
        \includegraphics[width=\textwidth]{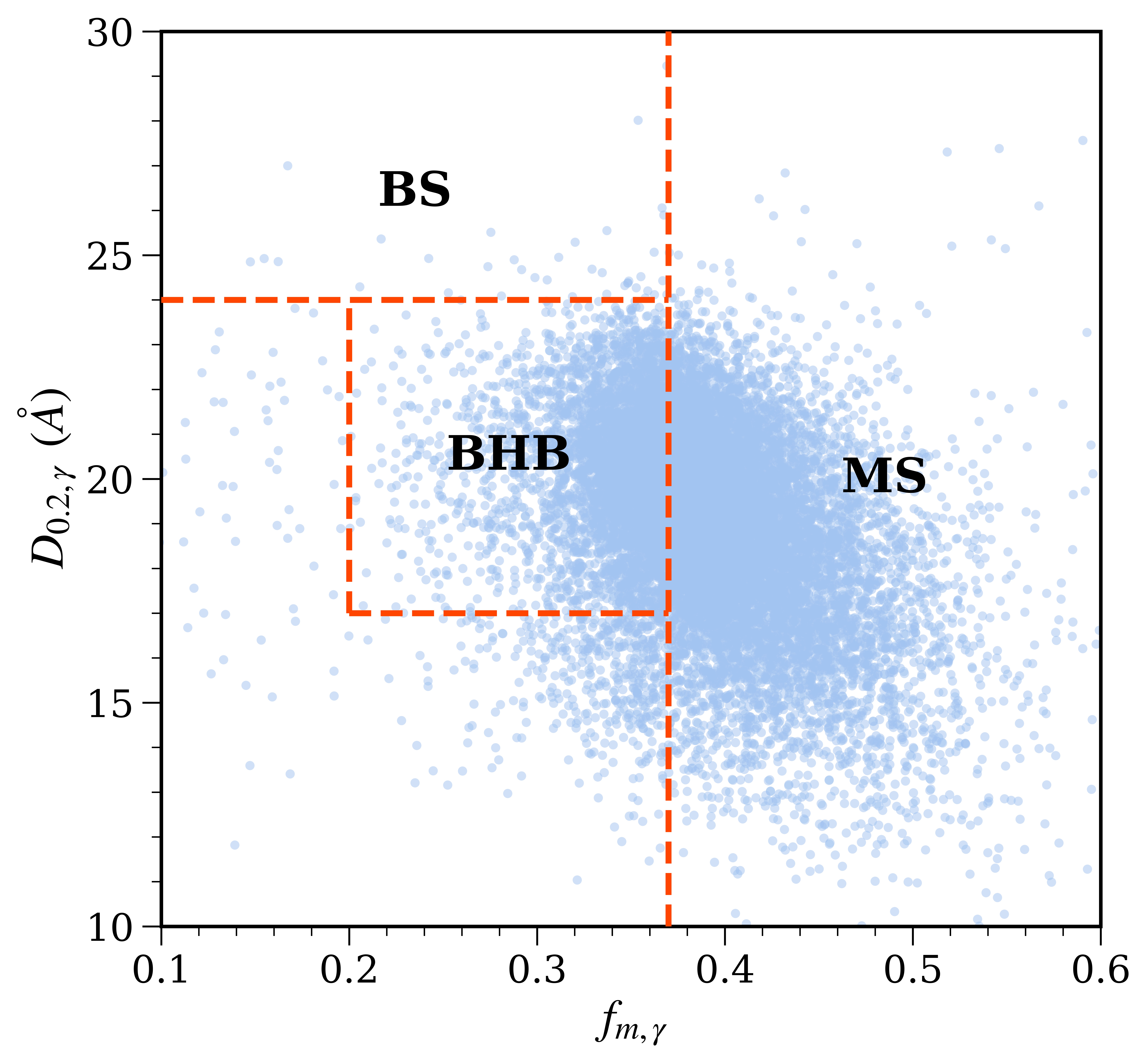} 
    \end{minipage}
    \caption{Distribution of BHB candidates in the $D_{0.2}$ vs. $f_m$ plane for H$\delta$ (left panel) and H$\gamma$ (right panel). The dashed lines indicate the selection boundaries adopted from \citet{ju2024}.}  
    \label{fig:Dfm_plot}
\end{figure}

Figure~\ref{fig:Dfm_plot} presents the distribution of $D_{0.2}$ and $f_m$ measurements, where the approximate regions occupied by BHB, BS, and MS stars are labeled. BS tends to show larger $D_{0.2}$, late-type MS star generally has higher $f_m$, and hot subdwarfs typically exhibit smaller $D_{0.2}$ with relatively large $f_m$. To identify the final sample, we applied the specific selection criteria defined by \citet{ju2024}:

\begin{equation}
\begin{array}{l}
18 \leq D_{0.2,\delta} \leq 25~\text{\AA}, \quad 0.2 \leq f_{m,\delta} \leq 0.36;\\
17 \leq D_{0.2,\gamma} \leq 24~\text{\AA}, \quad 0.2 \leq f_{m,\gamma} \leq 0.37.
\end{array}
\label{eq:selection_criteria}
\end{equation}

Among 16,721 spectra with available spectral coverage, 7542 and 4484 spectra satisfied the criteria for H$_{\delta}$ and H$_{\gamma}$, respectively. The intersection of these two selections yielded 3923 spectra corresponding to 3583 unique sources. We adopt these objects as the final BHB sample for subsequent analyses.

\subsection{Stellar Properties of the Final BHB Sample}

To characterize the final sample of 3583 BHB stars and perform a physical plausibility check of our classification, we estimated their atmospheric parameters and absolute magnitudes. These quantities enable a comparison with the canonical BHB loci in parameter space.

\subsubsection{Estimation of Atmospheric Parameters}

We estimate $T_{\mathrm{eff}}$, $\log g$, and [Fe/H] for our BHB stars using the Stellar Label Machine (SLAM; \citealt{ju2025}). SLAM incorporates multi-band color information during training to better constrain stellar temperatures. We therefore cross-match our sample with Gaia EDR3 \citep{gaiaedr3} to obtain $BP$, $G$, and $RP$ photometry, and with 2MASS \citep{skrutskie2006} to obtain $J$ and $H$ magnitudes for the common sources. Reddening corrections are applied using the 3D dust map of \citet{green2019}. We follow the implementation and preprocessing procedures described in \citet{ju2025}.

\begin{figure}[ht]
    \centering
    \includegraphics[width=0.75\textwidth]{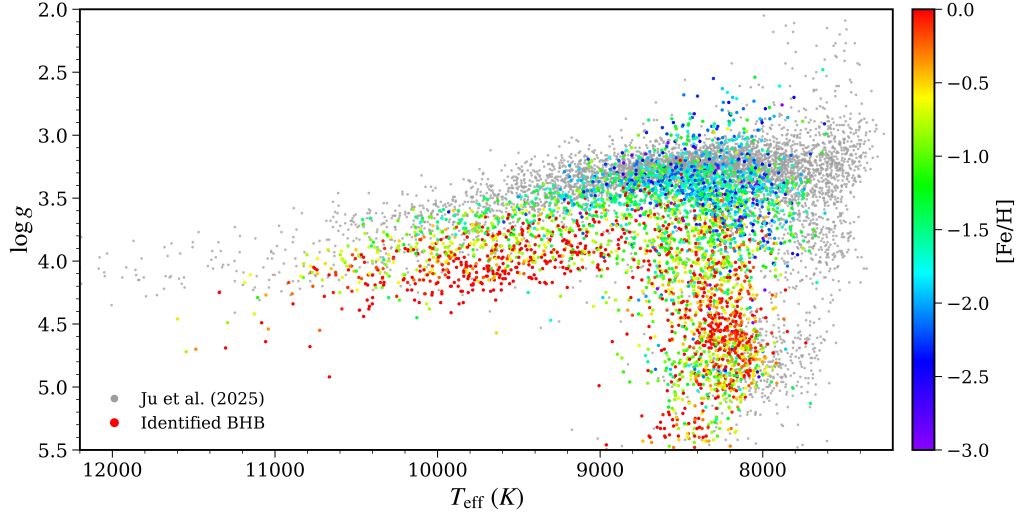}
    \caption{Distribution of BHB stars in the $T_{\mathrm{eff}}$--$\log g$ plane, color-coded by [Fe/H]. Gray dots show the literature BHB sample from \citet{ju2025}, while colored points highlight the 3583 newly identified BHB stars in this work.}
    \label{fig:atmos}
\end{figure}

As shown in Figure~\ref{fig:atmos}, our SLAM-based labels exhibit a distribution pattern broadly consistent with that reported by \citet{ju2025}, while a non-negligible fraction extends beyond the canonical parameter range expected for BHB stars. Among the 3583 identified BHB stars, 667 objects have $\log g>4.5$, forming a concentrated feature primarily at $T_{\mathrm{eff}}\approx 8000$--$9000$~K. In addition, 1941 stars show relatively high metallicities ([Fe/H]$\geq -1$); these metal-rich solutions are distributed over a broader temperature range and partially overlap with the high-$\log g$ subsample.

Following \citet{ju2025}, we interpret these apparent deviations mainly as inference-related systematics arising from low-quality or peculiar spectra and from limited training-set coverage that forces the model to extrapolate in a regime where the labels become more degenerate. Because $T_{\mathrm{eff}}$ is anchored by broadband colors (e.g., $BP-RP$), it is comparatively robust, whereas $\log g$ and [Fe/H] depend more strongly on detailed line profiles and thus are more vulnerable to such effects. We note that physical processes may still contribute at some level: atmospheric diffusion in hot BHB stars can enhance apparent metal abundances \citep{behr2003a}, and helium-related spectral peculiarities (as suggested for a minority of outliers by \citealt{ju2025}) could also affect the inferred [Fe/H] and/or $\log g$, but these are expected to be secondary compared to the dominant spectral/model limitations.

Nevertheless, we cannot exclude the possibility that a fraction of these objects are not genuine BHB stars. We list their inferred parameters in Table~\ref{tab:catalog} for future inspection and follow-up analyses.

\subsubsection{Absolute Magnitudes and Color-Magnitude Diagram}

\begin{figure}[ht]
    \centering
    \begin{minipage}{0.42\textwidth}
        \centering
        \includegraphics[width=\textwidth]{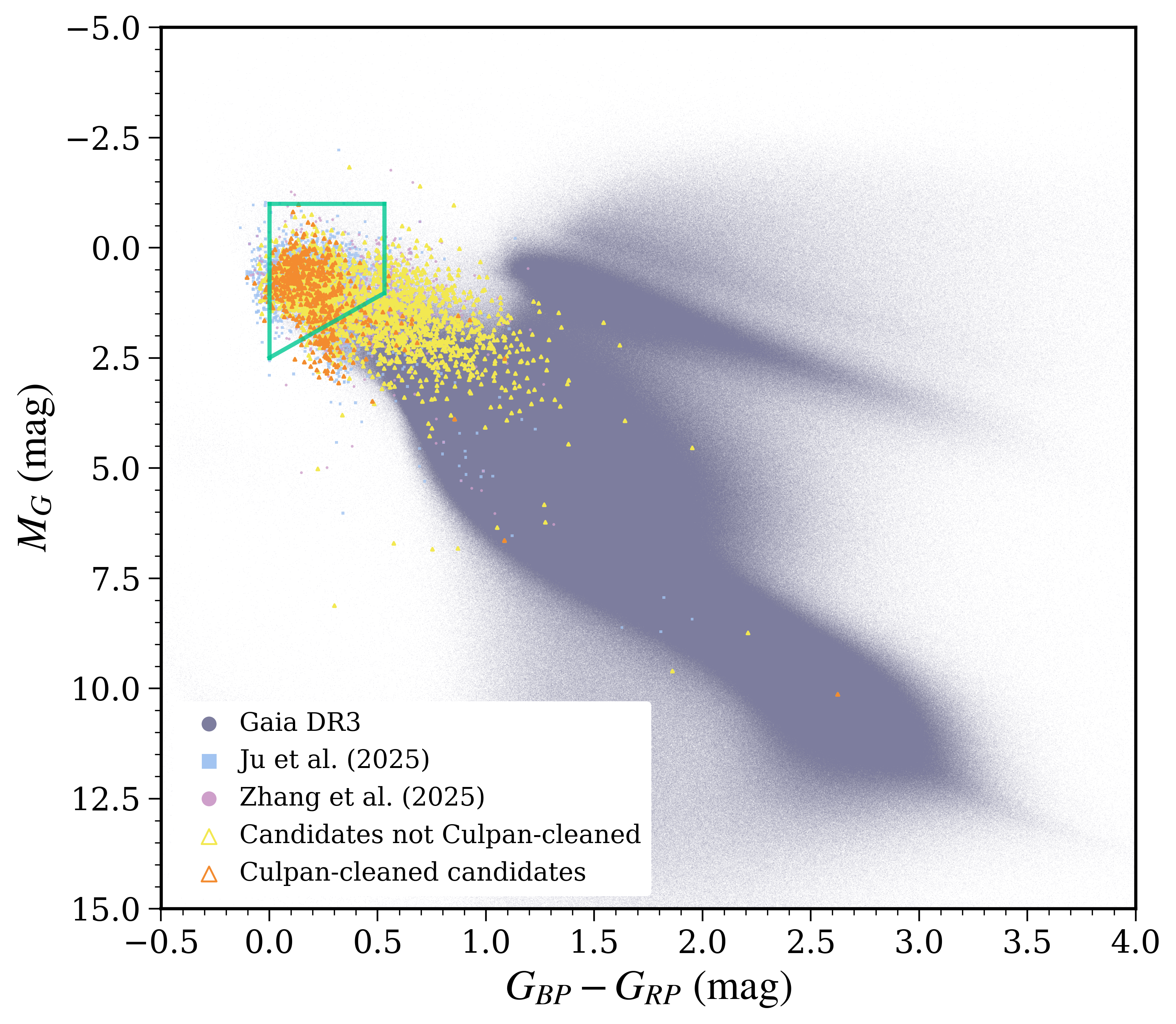} 
    \end{minipage}
    \begin{minipage}{0.42\textwidth}
        \centering
        \includegraphics[width=\textwidth]{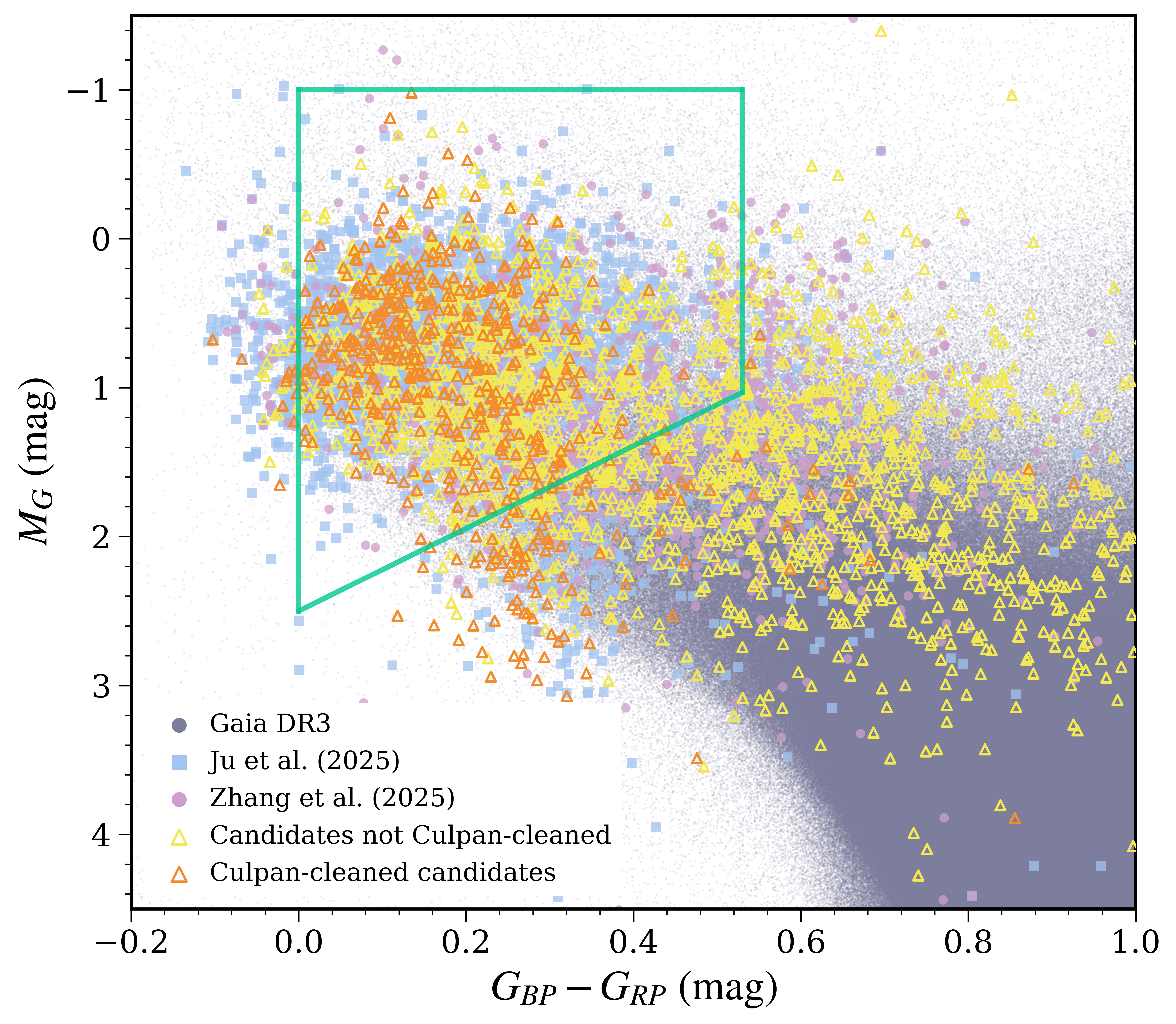} 
    \end{minipage}
    \caption{Color-magnitude diagrams (CMDs) constructed from Gaia DR3 data. In both panels, our BHB candidates are shown together with literature BHB samples from \citet{ju2025} and \citet{zhang2025}, overlaid on the Gaia source catalog (gray background). The left panel shows the full CMD, while the right panel zooms into the BHB regime. Open triangles denote the 2520 candidates identified in this work: orange open triangles indicate the subset that satisfies the spatial and kinematic criteria in Table~1 of \citet{cuplan2021} (used to mitigate contamination), whereas yellow open triangles mark the remaining candidates that do not meet these criteria. Blue squares represent the BHB sample from \citet{ju2025}, and purple circles represent the BHB sample from \citet{zhang2025}. The green trapezoid denotes the BHB selection boundary adopted from \citet{cuplan2021}.
}
    \label{fig:cmd}
\end{figure}

To assess the photometric consistency of our spectroscopically selected candidates, we constructed color–magnitude diagrams (CMDs) and compared their loci with those of established samples in the literature.
To derive absolute $G$-band magnitudes ($M_G$), we first cross-matched our initial sample of 3583 stars with the LAMOST DR12 catalog using their OBSIDs to retrieve the \texttt{gaia\_source\_id} field. Utilizing this identifier, which corresponds to the \texttt{source\_id} in Gaia DR3, we then cross-matched our sample with the Gaia DR3 catalog to obtain reliable parallax-based distances. We applied standard astrometric quality cuts, selecting only sources with positive parallaxes and relative uncertainties below 20\% ($\varpi/\sigma_{\varpi} > 5$; \citealt{geier2017, pelisoli2019}). This selection yielded a high-quality subset of 2520 stars (denoted as yellow and orange triangles in Figure~\ref{fig:cmd}). For this subset, we calculated $M_G$  using the standard relation:

\begin{equation}
M_G = G + 5 + 5 \log_{10} (\varpi / 1000),
\label{eq:MG}
\end{equation}

where $G$ is the apparent magnitude and $\varpi$ is the parallax in milliarcseconds.

Figure~\ref{fig:cmd} presents the Gaia DR3 CMD of our candidates together with the BHB catalogs of \citet{ju2025} and \citet{zhang2025}. The green trapezoid indicates the canonical BHB selection region adopted by \citet{cuplan2021}: $-0.1 < G_{BP}-G_{RP} < 0.53$ and $-1 < M_G < 2.5 - 2.77\,(G_{BP}-G_{RP})$. In both panels, our parallax-selected subsample (yellow and orange triangles) broadly follows the BHB locus and shows no obvious systematic offset in color or absolute magnitude relative to the literature samples. Nevertheless, only $\sim$54\% of our parallax-selected objects lie strictly within the trapezoid. The remaining dispersion---most notably the redward extension---is expected because we do not apply extinction corrections in this figure, and because lines of sight toward the Galactic plane and the Magellanic Clouds suffer from severe crowding and enhanced contamination by Population~I main-sequence A/B stars and blue stragglers, which can also degrade Gaia $G_{BP}$ and $G_{RP}$ photometry \citep{cuplan2021}. To mitigate these specific effects, we applied the sky-region and kinematic exclusions listed in Table~1 of \citet{cuplan2021}, yielding a refined subset of 577 stars (orange triangles). After this cleaning, about $\sim$80\% of the remaining objects lie within the selection boundary.

\subsection{Catalog of identified BHB Stars}

We present a comprehensive catalog of 3583 newly identified BHB stars, a subset of which is listed in Table~\ref{tab:catalog}. The catalog includes the LAMOST spectral identifiers (Obsid), equatorial coordinates (R.A. and Decl.), and the $g$-band S/N (S/NR$_{g}$) from LAMOST DR12. Photometric information from Gaia DR3 comprises the color index $G_{\rm BP}-G_{\rm RP}$ and the derived absolute magnitude $M_G$; both quantities are left undefined for sources without matched Gaia photometry or with relative parallax error exceeding 20\% and positive parallax. Atmospheric parameters ($T_{\rm eff}$, $\log g$, and $[{\rm Fe/H}]$) are estimated using the SLAM model \citep{ju2025}. This expanded BHB sample provides a valuable tracer for studies of the structure, formation, and dynamical evolution of the Milky Way. The source code used to identify these stars is publicly available at \url{https://github.com/Eric-star666/MSFA-Net}.

\begin{deluxetable*}{@{}l c c c c c c c c@{}}
    \setlength{\tabcolsep}{3.5mm}
    \tablecaption{The Catalog of BHB Stars in LAMOST DR12\label{tab:catalog}}
    \tablewidth{0pt}
    \tablehead{
        \multicolumn{1}{@{}l}{Obsid} &
        \colhead{R.A. (deg)} &
        \colhead{Decl. (deg)} &
        \colhead{S/NR$_{g}$} &
        \colhead{$G_{\rm BP}-G_{\rm RP}$} &
        \colhead{$M_G$} &
        \colhead{$T_{\rm eff}$ (K)} &
        \colhead{$\log g$} &
        \multicolumn{1}{c@{}}{$[{\rm Fe/H}]$}
    }
    \startdata
    1230412066 & 175.695220 &  7.789808 &  86.12 & 0.025963 & 0.044144 & 8982.60 & 3.31 & -1.95 \\
    1017411156 & 266.029558 & 27.898654 &  96.03 & 0.202767 & 0.610264 & 8373.87 & 3.46 & -1.23 \\
    842113049  &  21.347900 & 35.060278 & 219.41 & 0.247738 & 0.544068 & 8046.81 & 3.40 & -3.31 \\
    820307114  & 249.277837 & 26.200366 & 179.13 & 0.142282 & 1.074209 & 9129.97 & 3.56 & -1.21 \\
    679416161  &   3.906428 & 34.324456 & 271.78 & 0.155095 & 1.032125 & 8561.56 & 3.33 & -1.97 \\
    \enddata
    
    \tablenotetext{}{\hspace{-0.2 cm}(This table is available in its entirety in machine-readable form in the \href{https://zenodo.org/records/18437827}{online article}.)}

\end{deluxetable*}

\section{Summary} \label{sec:sum}

In this study, we propose MSFA-Net, a two-stage deep learning framework for identifying BHB stars from LAMOST low-resolution spectra. The model couples a hierarchical multi-scale residual backbone with an explicit FFT-based soft frequency attention module, enabling joint modeling of local spectral features and global frequency-domain context. Using transfer learning, MSFA-Net first performs five-class candidate screening and then refines the results with a fine-tuned binary classifier, achieving test-set precisions of 94.67\% and 98.07\%, respectively.

We applied MSFA-Net to LAMOST DR12 in a stepwise manner. The five-class model initially selected 37,441 candidate spectra, which were subsequently reduced to 27,853 spectra by the binary refinement stage. We then removed previously known objects and applied stringent Balmer-line profile criteria, yielding 3923 newly identified high-confidence BHB spectra. After merging duplicates, this corresponds to 3583 unique newly identified BHB stars.

To further characterize the sample, atmospheric parameters  ($T_{\mathrm{eff}}$, $\log g$, and [Fe/H]) were inferred using the machine-learning-based SLAM model. The overall parameter distribution is consistent with the expected BHB locus; however, a subset of stars exhibits elevated $\log g$ and relatively high [Fe/H]. We attribute these outliers primarily to inference-related systematics (e.g., limited training-set coverage and spectral quality/peculiarity) and report their inferred parameters for future inspection.

We also cross-matched our BHB catalog with Gaia DR3 to examine photometric colors and absolute magnitudes, providing an independent validation of consistency with the expected BHB sequence.

Overall, our results demonstrate the effectiveness of MSFA-Net for mining BHB stars from large-scale spectroscopic surveys. The resulting catalog of 3583 newly identified BHB stars provides a valuable legacy dataset for investigating halo substructure and constraining the assembly history of the Milky Way.

\section*{Acknowledgments}

This work was supported by the Natural Science Foundation of Shandong Province (grant No.~ZR2025MS06).

We thank the LAMOST team for providing the spectroscopic data used in this study. LAMOST (the Large Sky Area Multi-Object Fiber Spectroscopic Telescope; the Guoshoujing Telescope) is a National Major Scientific Project operated by the National Astronomical Observatories, Chinese Academy of Sciences, and funded by the National Development and Reform Commission. Additional information is available at \url{https://www.lamost.org/}.

\bibliographystyle{aasjournal}
\bibliography{sample631}{}



\end{document}